\documentstyle[11pt,epsf]{article}
\newcommand{\sptwo}{1.6}

\newcommand{\doublespace}{\edef\baselinestretch{\sptwo}\Large\normalsize}

\begin{document}

\doublespace

\begin{center}
{\large \bf The Double Quantum Dot Feline Cousin of Schr\"odinger's Cat:
An Experimental Testbed for a Discourse on Quantum Measurement Dichotomies}
\end{center}

\medskip

\begin{center}
{\bf S. Bandyopadhyay$\footnote{Corresponding author. E-mail: bandy@quantum1.unl.edu}$}\\
{\it Department of Electrical Engineering,
University of Nebraska,
Lincoln, Nebraska 68588-0511, USA}
\end{center}

\medskip

\begin{abstract}
\noindent Quantum measurement theory is a perplexing discipline 
fraught with paradoxes and dichotomies. Here we discuss a
gedanken experiment that uses a popular testbed - namely, a coupled double quantum dot system - to revisit intriguing
questions about the collapse of wavefunctions, irreversibility,
objective 
reality 
and the actualization of a measurement outcome.
\end{abstract}

\twocolumn

Quantum measurement theory is a sub discipline replete with many 
subtleties of 
quantum mechanics. Its basic underpinning can be summarized by
a fundamental and yet profound question: when and 
how does a pure state, descriptive of a quantum
system entangled with a measuring apparatus (also a quantum system), 
evolve into a 
mixed state that results in distinguishable outcomes of the 
measurement. Since in standard quantum mechanics, 
 no unitary time evolution can cause a pure state to 
evolve into a mixed state there is essentially no cookbook
``quantum recipe''' to  forge  distinguishable 
outcomes \cite{home_book, weinberg}in quantum measurement.

A number of formalisms that augment the standard mathematical 
framework of quantum mechanics \cite{ghirardi, penrose} provide
a dynamical description of the measurement process in 
terms of an actual transition of a pure state into a 
mixed state. This has been termed ``collapse of a 
wave function''. However, even if we accept the 
augmented mathematical framework,
some mysteries still remain.
How does the collapse occur? Is it a discrete event in
time or is it a continuous process? Is the collapse observer-dependent
(i.e. it happens only when an observer decides to look at the 
outcome of a quantum measurement) or does the outcome materialize at some 
time independent of the observer? In this short communication,
we re-visit these  issues in the context of a 
popular quantum system that illustrates many of the subtleties 
in quantum measurement theory.

Consider a  double quantum dot system coupled  by a translucent 
tunnel barrier.
The conduction band diagram is shown in Fig. 1(a).
The two quantum dot materials are identical in all respects 
except in their elastic constants. That is, electrons cannot
distinguish between them, but {\it phonons can}.
\begin{figure*}
\epsfxsize=5.8in
\epsfysize=2.4in
\centerline{\epsffile{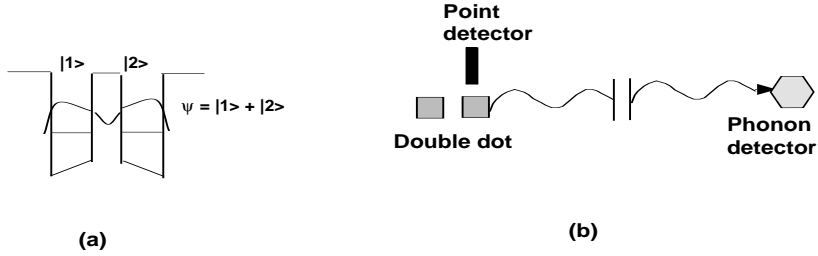}}
\caption{(a) The conduction band profile of two semiconductor 
quantum dots with an intervening tunnel barrier. All subbands 
are aligned in energy to allow resonant tunneling of electrons
between the two dots. The only difference in the material 
of the two dots is in their elastic constants. (b) The experimental 
set-up.}
\end{figure*}
An electron is introduced into the ground state of the system
and exists in a coherent superposition of two states $|1>$ and $|2>$
\begin{equation}
\psi = {{1}\over{\sqrt{2}}}( |1> + |2> )
\end{equation}
where $|1>$ is a semi-localized wave function in the left dot
and $|2>$ is a semi-localized wave function in the right dot.
A weakly coupled point detector in the vicinity
of one of the dots can tell whether  that 
dot is occupied by the electron or the other one is.
This experimentally realizable system has been studied in the 
context of the quantum 
measurement problem by a number of authors \cite{gurvitz,
korotkov, stodolsky} recently.

We now summarize three different viewpoints regarding the 
quantum measurement problem. The orthodox viewpoint 
associated with the Copenhagen interpretation is epitomized
by  Von-Neumann:
the wave function collapses  when an observer chooses to look at the 
detector and gain knowledge about where the electron is \cite{von-neumann}.
This is an observer-dependent reality and has been 
much discussed in the context of the Schr\"odinger cat paradox. A different viewpoint  
espoused by a number of researchers \cite{bohm, gisin, omnes,
ghirardi} is predicated on objective reality. It can be briefly stated as follows: once a measurement
outcome is actualized, it remains ``out there'' forever to 
be inspected by an observer at {\it any} subsequent time
without changing the outcome. The outcome does not depend 
on when, or if at all, the observer inspects it, and does 
not change once actualized. Home and Chattopadhyay \cite{home} have
suggested an experiment involving UV-exposed DNA molecules
to empirically determine at what {\it instant} an outcome is actualized
and the result recorded in a stable and discernible form for perpetuity.  
A third viewpoint \cite{gurvitz} claims that there may be no
such precise {\it instant}. The pure state  may {\it
gradually} evolve towards a mixed state and concomitantly
decoherence begins 
to set in, but the system may
never quite completely decohere in a finite time (we define complete decoherence
as the state in which the off-diagonal terms of the 2$\times$2 density matrix
associated with Equation (1) vanish). 
The off-diagonal terms may decay with time owing
to the interaction with the detector (and this may slow
down the {\it wiederkehr} quantum oscillation between the states $|1>$ 
and $|2>$ 
- the so-called quantum Zeno effect) but the 
off-diagonal terms need not {\it ever} vanish completely. This has been
termed a ``continuous collapse''. Korotkov
\cite{korotkov} claims that continuous measurement need not
cause {\it any} decoherence or collapse (i.e, the off-diagonal
terms need not decay at all because of the interaction
with the detector) if continuous knowledge 
of the measurement result at all stages of detection is used 
to faithfully reconstruct the pure state. These three viewpoints
are quite disparate and cannot be reconciled easily.

We suggest a simple gedanken experiment to resolve some of
 these conflicting viewpoints. Consider
the situation when we have two independent
detectors capable of detecting which dot is occupied by the 
electron in Fig. 1. The detectors are independent in the 
sense that they are located vast distances apart and initially there
is no coupling between them. One detector is the weakly
coupled point detector (see Fig. 1b) in the vicinity of a dot capable of 
fairly non-invasive measurement which causes at most
gradual collapse a l\'a Gurvitz. The other detector is 
a phonon detector located far away. Suppose that when the
electron is in the right dot it emits a {\it zero energy}
acoustic phonon which has a finite wave vector and hence a finite
momentum. It also has a finite group velocity.
Such phonons do not typically exist in bulk 
materials, but exist in quantum confined structures 
like wires \cite{svizhenko} and dots. The emitted phonon
has  different wave vectors depending on whether the 
emission took place in the left dot or the right dot
because elastic constants (and hence the phonon dispersion
relations) in the two dots are different. When the phonon
arrives at the detector, it is absorbed by an electron
and by measuring the momentum imparted to the electron
(or equivalently the associated current), one can tell whether the
phonon came from the left dot or the right dot. Thus, monitoring
the current in the phonon detector will constitute a ``measurement''.
Let us say that the phonon was emitted at time $t$ = 0$\footnote{
It may bother the reader that Heisenberg's Uncertainty Principle
is being violated in this thought experiment. If the phonon
has precisely zero energy, how can we say that it is emitted
at exactly time t=0? The answer is that at time t=0, we are 
not {\it measuring} the energy. If we ever wanted to measure the
phonon's energy, we could take forever. If indeed Heisenberg's
Uncertainty Principle were relevant here, then {\it all}
elastic collisions (e.g. electron-impurity collision) will
take forever. Yet we can calculate an effective scattering
time for an electron impurity collision from Fermi's
Golden Rule.}$ and it arrives at the phonon detector at 
time $t$ = $t_1$. The detector
finds that the phonon came from the right dot.

If the viewpoint of objective reality \cite{home, bohm, gisin, omnes,
ghirardi} is correct, then the actualization of the outcome took 
place at time $t$ = 0. Thereafter, the electron will be 
always found in the right dot.
We can empirically
pinpoint this instant at a later time
$t$ $>$ 0 (actually at $t$ $\geq$ $t_1$)
since we can determine $t_1$, the time of flight of the 
phonon between the dot and the phonon detector. We simply have to know
the distance between the dot and the detector and the phonon group 
velocity to know $t_1$. Thus when the phonon detector registers the phonon,
we will know that the actualization took place $t_1$ units of 
time prior to the registration event.
Additionally, if we know the time $t$ = -$t_2$ when the 
electron was injected into the double dot system, then 
we can find out how long thereafter the actualization of 
the outcome took place (this time is simply $t_2$). This is
similar to what Home and Chattopadhyay had proposed to achieve
in their UV-exposed DNA system \cite{home}.

We now come to the central issue. Between the time $t$ = 0 and 
$t$ = $t_1$ (i.e. while the phonon is in flight), the observer
(phonon detector) is still ignorant of the outcome,
but the actualization of the measurement \cite{home}
has supposedly already taken place. During this critical 
time period, the
weakly coupled point detector tries to 
{\it continuously} determine which dot is occupied. 
If the {\it observer-independent} viewpoint is correct,
then the electron will be always found in the right dot. 
 But, if the {\it observer-dependent} 
viewpoint is correct \cite{von-neumann}, then the 
Schr\"odinger cat is  in suspended animation between 
$t$ = 0 and $t$ = $t_1$ since the observer (phonon 
detector) has not registered any phonon yet. Consequently,
 the almost non-invasive point detector (which takes a very long
time to destroy the superposition acting alone) should have a
non-zero probability of finding the electron in the left dot.
To ensure that these are the only two possible scenarios, we will
allow the maximum latitude.
For instance, we will assume: (i) the quantum oscillation
period between the two dots ({\it wiederkehr}) is much smaller than the time of flight $t_1$
and the Zeno effect \cite{misra} is negligible because of the weak 
coupling with the non-invasive point detector,
(ii) the emission of zero energy phonon does not alter the electron's
energy and hence does not subsequently disallow resonant tunneling between 
the quantum dots, and (iii) the remote phonon detector is unaware of the set-up
before time $t$ = $t_1$ and hence cannot influence events before 
time $t$ = $t_1$ (causality). Thus, if the point detector ever finds the 
electron in the left dot between $t$ = 0 and $t$ = $t_1$, the 
objective reality (observer-independent) viewpoint will be suspect.
In this pathological example, the difference between the observer-dependent 
and observer-independent viewpoint can be simply stated thus:
in the first viewpoint, the collapse took place at $t$ = $t_1$ and in
the second viewpoint, it took place at $t$ = 0. As long as any non-invasive
detector in the timeframe $t$ = 0 till $t$ = $t_1$ finds the electron
in the left dot and the phonon detector at time $t_1$ 
finds the electron to have 
emitted the phonon in the right dot, we will know that the ``collapse'' 
did not take place at $t$ = 0 which would then contradict the 
observer independent viewpoint. We will then be forced to admit 
that perhaps collapse ultimately takes place in the sensory
perception of the observer \cite{aicardi}. This is currently
a contentious topic.

An interesting question is whether the phonon 
emission is a collapse event. There is no energy 
dissipation involved in emitting a zero-energy phonon,
but energy dissipation is not necesssary for collapse since 
{\it elastic} interaction of an electron with a magnetic 
impurity that causes a change in the internal degree of 
freedom of the scatterer (say, spin flip) constitutes 
effective collapse. ``Creation'' of a phonon is certainly changing
its internal degrees of freedom in a major way and therefore
should be viewed as a collapse event within the framework
of standard models.

But what if the point detector will find the 
electron in the left dot {\it after} time $t$ = $t_1$ when 
the phonon detector has already determined that the electron
collapsed in the right dot. This will
make standard collapse models suspect \cite{leggett} since 
we must then admit that the phonon emission did not cause a collapse.
 Complete collapse is an irreversible 
event (equivalent to saying that the Zeno time is infinite).
However the third viewpoint of Gurvitz \cite{gurvitz}
guarantees that the electron will be ultimately delocalized
(and hence found in the left dot with a non-zero
probability) if we make a continuous measurement with the point detector.
In contrast, if frequent repeated measurements are made, then the Zeno
effect guarantees that the opposite will happen; the electron
will become more localized in one dot as the frequency of 
observation is increased. Thus, there is an essential 
dichotomy when one considers the fact that a continuous measurement
is really the ultimate limit of  frequent repeated measurements
and yet they make opposite predictions. It is 
not clear how this dichotomy will be ultimately resolved. 

In this communication, we have proposed a gedanken experiment 
to resolve some of the \\
dichotomies between the myriad viewpoints
permeating quantum measurement theory. Experiments such as the one 
proposed here
will soon be within the reach of modern technology. Hopefully,
they will shed new light on this fascinating topic.

\bigskip


\end{document}